\documentclass[12pt,dvips]{article}

\usepackage{amsmath,amssymb,exscale}
\usepackage{array,multicol}
\usepackage{afterpage,float,flafter}
\usepackage{epsfig}
\def\beq{\begin{equation}}
\def\beqn{\begin{eqnarray}}
\def\eeq{\end{equation}}
\def\eeqn{\end{eqnarray}}
\def\lp{\left}
\def\rp{\right}
\@addtoreset{equation}{section} \makeatother

\title{
\vspace{1cm}
\Large\textbf{Consistent Lorentz Violation in Flat and Curved
Space}
\vspace*{.5cm}
\author{
\large \textbf{Gia Dvali,\footnote{email: gd23@nyu.edu}~~ Oriol Pujol{\` a}s\footnote{email: pujolas@ccpp.nyu.edu}~~and Michele Redi\footnote{email: redi@physics.nyu.edu}}\\
\emph{Center for Cosmology and Particle Physics,}\\\emph{Department of Physics, New York University}\\
\emph{4 Washington Place, New York, NY 10003}}}

\date{}
\begin{document}
\maketitle \thispagestyle{empty} \vspace*{.5cm}

\begin{abstract}

Motivated by the severity of the bounds on Lorentz violation in
the presence of ordinary gravity, we study frameworks in which
Lorentz violation does not affect the spacetime geometry. We show
that there are at least two inequivalent classes of spontaneous
Lorentz breaking that even in the presence of gravity result in
Minkowski space. The first one generically corresponds to the
condensation of tensor fields with tachyonic mass which in turn is
related to ghost-condensation. In the second class, realized by
the DGP model or theories of massive gravitons, spontaneous
Lorentz breaking is induced by the expectation value of sources.
The generalization to de-Sitter space is also discussed.

\end{abstract}

\newpage
\renewcommand{\thepage}{\arabic{page}}
\setcounter{page}{1}

\section{Introduction}

The possibility of breaking Lorentz invariance is a deep question.
In General Relativity, the continuous spacetime symmetries, such
as the Lorentz symmetry, are part of gauge invariance, and as
such, their breaking can only be understood as the low energy
limit of some underlying spontaneous breaking.

Most phenomenological studies in the literature consider explicit
breakings, in the limit where gravity is decoupled. The latter
however is expected to have very profound consequences, because
any order parameter that spontaneously breaks Lorentz invariance
will couple to gravity, at least minimally, distorting the
geometry of spacetime. Enforcing the cosmological bounds could
place such severe constraints on the order parameters, that the
corresponding Lorentz violation would have limited observational
interest.

This issue can be illustrated considering the simplest possible
model of spontaneous Lorentz symmetry breaking, obtained through a
time dependent scalar field, $\phi$. Due to the  vacuum
expectation value (VEV), $\partial_0 \phi\neq0$, the Lorentz
symmetry is broken to the rotation subgroup and this breaking will
be communicated to ordinary particles (e.g., electrons or photons)
by higher dimensional operators of the form,
$$
{1 \over \Lambda_1^4}\partial_{\mu}\phi\partial_{\nu}\phi
\,\bar{\psi}_e\gamma^{\mu}\partial^{\nu}\psi_e +{1 \over
\Lambda_2^4}\partial_{\mu}\phi\partial_{\nu}\phi\, F^{\mu}_\alpha
F^{\nu\alpha}~,
$$
where $\Lambda_{1,2}$ are the scales of new physics. Substituting
the non-zero VEV of the scalar time derivative, the above
operators generate an effective Lorentz violating dispersion
relation for the electron and the photon, which could potentially
be observed if the corrections are at least of order $10^{-15}$
and $10^{-23}$ respectively \cite{coleman}.

However, taking into the account gravity, immediately makes the
prospect of such observations pretty dim. Indeed, the time
dependence of $\phi$ will also contribute to the energy density of
the Universe, which should not exceed the critical one because it
cannot be cancelled by a readjustment of the cosmological constant
or any other conventional source. This implies that at best
$\partial_0\phi \sim (10^{-3}eV)^2$. Demanding that the Lorentz
violating contributions to the dispersion relation are potentially
observable without conflicting with the cosmological bound
requires that $\Lambda_1 \lesssim 10\, eV$ and $\Lambda_2 \lesssim
KeV$~! Such small values are ruled out by particle physics
observations. The situation is a little better when the Lorentz
violation affects the mass parameters. For example for the photon,
the operator
$$
{1\over\Lambda_3^2}\partial_{\mu}\phi\partial_{\nu}\phi\,
A^{\mu}A^{\nu}~,
$$
with $\Lambda_3$ as high as $TeV$ \cite{papucci}, would still
produce Lorentz violating effects not far from the observable
bounds.

Thus, it is very important to understand the consistent ways of
spontaneous  breaking of the Lorentz invariance, which even in the
presence of gravity do not affect the background geometry. For
instance, having Lorentz violation in exact Minkowski or de Sitter
spaces. Such theories would provide a consistent underlying
framework for the phenomenological studies of Lorentz violations
such as \cite{coleman}.

In this note we investigate different scenarios of such Lorentz
violation. A large class of Lorentz violating theories which do
not change the background corresponds to the condensation of
tachyons of arbitrary integer spin fields. We show that these
theories in a certain limit of parameter space effectively become
equivalent to some type of ghost condensation \cite{arkani}. The
basic feature of these theories is that nowhere in parameter space
one can recover a consistent Lorentz invariant vacuum. This is
different from the usual breaking of internal symmetries where
instead there is a consistent effective field theory description
of the unbroken phase. The reason for this is that the instability
that breaks Lorentz invariance can only be stabilized by cutoff
sensitive physics. While this may cast some shadows on the
possibility to find ultraviolet completions of these theories,
fluctuations around the Lorentz breaking vacuum can give rise to
consistent low energy effective theories.

We point out that there exists a second inequivalent class of
theories in which spontaneous Lorentz breaking is accomplished due
to the VEV of the energy momentum tensor. This breaking is analogous
to the one induced by a cosmological fluid however the space remains
Minkowskian and time independent. We show that this possibility is
in fact realized in the DGP model \cite{dgp} and massive gravity
theories. Contrary to the previous case there is no strong
sensitivity to the cut-off and there is a smooth limit in parameter
space where a Lorentz invariant ground state is recovered. Similarly
we show that in de-Sitter space it is possible to break the
isometries of the space while keeping the space maximally symmetric.
In this case however the breaking necessarily dilutes with the
expansion of the Universe.

\section{High Spin Tachyons}
\label{hs}

A natural way to try to spontaneously break Lorentz invariance in
flat space is by following the analogy with the breaking of internal
continuous symmetries. For this purpose one needs to introduce a
bosonic integer spin order parameter $A_{\mu_1,..\mu_n}$ and make it
condense. Since $A$ transforms non-trivially under Lorentz, its VEV
spontaneously breaks Lorentz invariance. In the presence of gravity
the cosmological constant in the action can be chosen so that the
ground state remains exactly Minkowski space. In order to achieve
this, one needs two ingredients. Schematically, a tachyonic mass
term,
\begin{equation}
-m^2 A^2
\end{equation}
and a stabilizing (non-derivative) self-interaction, e.g.,
\begin{equation}
\lambda\, A^4
\end{equation}
The seeming exception is a spin-0 field for which what needs to
condense is actually the derivative of a ghost-field
\cite{arkani}. However this difference is only naive. The reason
for this is that for higher spin fields the tachyonic and ghost
instabilities are inseparable. To illustrate the point let us
consider first a massive spin-1 field (see also \cite{luty} and
references therein),
\begin{equation}
S=\int d^4 x\, \left(-\frac 1 4 F^2- \frac {m^2} 2 A^2\right)
\end{equation}
The gauge invariance of the action can be restored introducing a
St\"uckelberg field that compensates the variation of the gauge
field under a gauge transformation. The action becomes,
\begin{equation}
S=\int d^4 x\, \left( -\frac 1 4 F^2 - \frac {m^2} 2 (\partial \phi+A)^2
\right)
\end{equation}
which is manifestly invariant under,
\begin{eqnarray}
\delta A_\mu&=& \partial_{\mu} \epsilon \nonumber \\
\delta \phi&=& -\epsilon.
\end{eqnarray}
It is clear (we use mostly plus signature) that a tachyonic mass for
the gauge field ($m^2<0$) is associated with a ghost-like kinetic
term for the St\"uckelberg field. This is the tachyonic counterpart
of the standard Higgs effect where a massless spin-1 field acquires
mass by eating a Goldstone boson with positive kinetic term.

The mechanism outlined above is actually completely general. Take
for simplicity a massive spin-2 field (up to some technical
modification the same holds for higher spin, see appendix A). The only
consistent mass term for the graviton around Minkowski vacuum is
described by the Pauli-Fierz (PF) combination,
\begin{equation}
V=\frac {m^2\,M_4^2} 8 (h_{\mu\nu}^2- h^2), \label{pf}
\end{equation}
where $h_{\mu\nu}=g_{\mu\nu}-\eta_{\mu\nu}$. The gauge invariance
of the mass term can be restored introducing St\"uckelberg fields,
\begin{equation}
h_{\mu\nu}=\hat{h}_{\mu\nu}+\partial_{\mu}
A_{\nu}+\partial_{\nu}A_{\mu}. \label{decomposition}
\end{equation}
The mass term is then invariant with respect to,
\begin{eqnarray}
\delta \hat{h}_{\mu\nu}&=&
\partial_{\mu}\xi_{\nu}+\partial_{\nu}\xi_{\mu}\nonumber \nonumber \\
\delta A_\mu&=&-\xi_{\mu}\nonumber \label{2variation}
\end{eqnarray}
By plugging (\ref{decomposition}) into the PF mass term one can
see that $A_\mu$ has the kinetic term of a gauge field. The sign
is such that a tachyonic instability for the spin-2 is translated
into a ghost instability for the St\"uckelberg field. Actually
$A_\mu$ has three degrees of freedom, which can be split into
spin-1 and spin-0 polarizations. Since the spin-0 component mixes
with $\hat{h}_{\mu\nu}$ it is useful to diagonalize the kinetic
terms. This can be achieved through the decomposition,
\begin{equation}
h_{\mu\nu}=\tilde{h}_{\mu\nu}+\partial_{\mu} \tilde
A_{\nu}+\partial_{\nu}\tilde A_{\mu}+\frac 1 {3 m^2}
\partial_\mu\partial_\nu \phi +\frac 1 6 \eta_{\mu\nu} \phi.
\label{decomposition2}
\end{equation}
With this, the scalar St\"uckelberg acquires a positive kinetic
term even for $m^2<0$. This can be understood from the fact that a
higher spin field becomes massive by eating a "massive" ghost
which in turn can be decomposed into massless ghost and a tachyon
St\"uckelberg with positive kinetic term.

We should here mention that although the mass term for ghost-free
theories is uniquely defined, it becomes unclear what form should
be used if one wants to create instabilities. The guideline in
this case is that if one does not want to increase the number of
propagating degrees of freedom, then one has to use the form
(\ref{pf}) even for the tachyonic case. Of course, such theories
can only make sense if the tachyonic instability is stabilized by
some non-linear self-interaction\footnote{In fact, such self
interactions could be used to stabilize extra time dimensions
\cite{extratime}, since these theories reduce to a tower of spin-2
tachyons (see Appendix A).}. While it is not the purpose of this
work to discover the form of such self-interaction, one natural
choice would be to take the stabilizing potential to be a function
of the PF combination. Our main point however is that any viable
model of spontaneous Lorentz breaking through the condensation of
integer spin tachyons, will be accompanied by ghost type
instabilities, which have to be stabilized in the true vacuum of
the theory, if such exists. As a result these types of theories
cannot have a consistent Lorentz invariant vacuum because, in
order to recover such a vacuum by continuum change of parameters
one has to reverse the sign of the kinetic term of the condensing
ghost which requires going through infinite strong coupling. In
other words the ghost instabilities are entirely dominated by
cut-off sensitive physics.

\section{Lorentz Violation in DGP}
\label{dgp}

We now turn to the DGP model \cite{dgp}. This scenario is
specified by the action,
\begin{equation}
S= -\frac {M_4^2} 2 \int d^4x \,\sqrt{-g}  R_4- \frac {M_5^3} 2 \int d^4x
dy\, \sqrt{-G} R_5
\end{equation}
which describes a tensionless brane with induced kinetic term for
the graviton embedded in a five dimensional empty bulk. (Later, we
shall consider the generalization with brane and bulk cosmological
constants.)

In DGP there are known static solutions that break Lorentz
invariance in Minkowski space \cite{lue}. By taking an energy-momentum
tensor localized on the brane of the form,
\begin{equation}
T_{MN}=\rho_0 \,\delta(y)\, {\rm diag} (0,-1,-1,-1,0) \label{emt}
\end{equation}
it is easy to check that the exact solution for the metric is
given by,
\begin{equation}
ds^2=-(1+ c |y|)^2 dt^2+ dx_1^2+dx_2^2+ dx_3^2 +dy^2
\label{metric}
\end{equation}
where $c=-\rho_0/M_5^3$. Moreover, in \cite{lue} it was shown that
the evolution of pressureless dust on a brane with negative
tension ($\rho_0<0$) automatically relaxes to such a ground state.

An important feature of the solution above is that it does not
depend on $M_4$ so that the DGP kinetic term is only needed to
reproduce $4D$ gravity at distances shorter than the crossover scale
$r_c=M_4^2/(2 M_5^3)$. Each slice at fixed $y$ is just Minkowski
space, however, due to the $y$ dependence, the solution globally
breaks Lorentz invariance to the rotation subgroup (in the preferred
frame) since the speed of light varies along the fifth dimension.
An observer living at fixed $y$ will
only detect the Lorentz violation through the non-relativistic
dispersion of bulk gravity modes\footnote{Lorentz violation will
also be transmitted at tree level to the Standard Model fields if
the brane has some thickness in the bulk or by couplings of the the
extrinsic curvature to Standard Model bilinears induced at loop
level.}.

The metric above is actually closely related to the domain-wall
solutions studied recently in \cite{oriol}. In that case it was
found that for a domain wall the metric is just the metric of a
codimension two object in $5D$, i.e. it is independent of the DGP
kinetic term. This similarity is no accident: the energy momentum
tensor (\ref{emt}) can be thought as a homogeneous distribution of
parallel (euclidean) domain walls. Since for codimension two
objects one can find exact solutions for an arbitrary number of
defects, the metric (\ref{metric}) can directly be obtained from
the domain wall one.

In the light of the above one can also construct space-times with
different Lorentz breaking patterns in Minkowski space. Consider a
distribution of Lorentzian domain walls,
\begin{equation}
T_{MN}=\rho_0 \,\delta(y)\, {\rm diag} (1,-1,-1,~0,~0)
\label{emt2}
\end{equation}
The metric is then given by,
\begin{equation}
ds^2=-dt^2+ dx_1^2 +dx_2^2+ (1+c |y|)^2 dx_3^2 +dy^2
\label{metric2}
\end{equation}
which breaks Lorentz invariance to the subgroup generated by
$(J_3, K_1, K_2)$. Note that for domain walls with positive
tension $c<0$, so the requirement that the space is regular at
$y=-1/c$ implies that $x_3$ is a periodic coordinate.

\section{Massive Gravity and Other Solutions}

Given the existence of Lorentz violating solutions in DGP it is
natural to wonder if they are equivalent to some kind of tachyon
condensation. That this is not the case can be already understood
from the fact that we can smoothly recover a Lorentz invariant
ground state by taking the source to zero. To clarify how this mechanism
works from the $4D$ point of view it is useful to deconstruct the DGP
solution of the previous section in terms of massive gravitons. This
will also show that this mechanism of Lorentz violation is general
to theories of massive gravitons.

To linear order consistent theories of massive gravitons are
governed by the PF equation,
\begin{equation}
{\cal E}^{\alpha\beta}_{\mu\nu} h_{\alpha\beta}+\frac {m^2} 2
(h_{\mu\nu}-\eta_{\mu\nu} h)=\frac {T_{\mu\nu}} {M_4^2},
\label{pfequation}
\end{equation}
where ${\cal E}^{\alpha\beta}_{\mu\nu}$ is the linearized Einstein
tensor,
\begin{equation}
{\cal E}^{\alpha\beta}_{\mu\nu} h_{\alpha\beta}=\frac 1 2 (\partial_\alpha \partial_\nu h^\alpha_\mu+\partial_\alpha \partial_\mu h^\alpha_\nu
-\partial_\mu\partial_\nu h-\square h_{\mu\nu}-\eta_{\mu\nu} \partial_\alpha\partial_\beta  h^{\alpha\beta}+\eta_{\mu\nu} \square h)
\end{equation}
In absence of sources the PF mass term guarantees that there are
no ghosts or tachyons and the vacuum is Lorentz invariant. Let us
now consider a constant energy momentum tensor.  The system admits
a static solution,
\begin{equation}
h_{\mu\nu}=-\frac {2} {m^2 M_4^2}\left(T_{\mu\nu}-\frac 1 3
\eta_{\mu\nu} T\right). \label{puregauge}
\end{equation}
This solution is pure gauge
($h_{\mu\nu}=\partial_\mu\xi_\nu+\partial_\nu\xi_\mu$) so the
space remains flat. However due to this VEV, probe particles will
detect a Lorentz breaking background  which is static and
Minkowskian. Of course whether the space remains flat to non
linear order will depend on the interaction terms.

We can now deconstruct the DGP solution. Consider fluctuations of
the metric around the vacuum,
$g_{\mu\nu}=\eta_{\mu\nu}+h_{\mu\nu}(x,y)$. At the linear level, DGP
reduces to a continuum of massive gravitons,
\begin{equation}
h_{\mu\nu}(x,y)=\int_0^{\infty} dm\,
h_{\mu\nu}^{(m)}(x)\,\psi^{(m)}(y)
\label{superposition}
\end{equation}
where the wave-functions $\psi^{(m)}(y)$ are determined by,
\begin{equation}
\left(\partial_y^2 + m^2+ m^2 r_c \delta(y)\right)
\psi^{(m)}(y)=0~, \label{waveequation}
\end{equation}
and each massive graviton satisfies the PF equation,
\begin{equation}
({\cal E} h^{(m)})_{\mu\nu}+ \frac {m^2} 2
(h_{\mu\nu}^{(m)}-\eta_{\mu\nu} h^{(m)})=\frac 1 {M_5^3} \int dy\,
\psi^{(m)}(y) T_{\mu\nu}(y). \label{kkgraviton}
\end{equation}
For the energy-momentum tensor (\ref{emt}) one gets,
\begin{equation}
h_{\mu\nu}^{(m)}=-\frac {2 \,\rho_0} {m^2 M_5^3}\,\psi^{(m)}(0)\,
{\rm diag}\,(1,~0,~0,~0)
\end{equation}
from which plugging in eq. (\ref{superposition}) one can reconstruct the
bulk solution ({\ref{metric}) to linear order.

The linear analysis also suggests that it might be possible to find
other Lorentz violating solutions in Minkowski space when $T_{00}$
is not zero. This simply follows from the fact that for any
$T_{\mu\nu}$ which is constant the massive graviton has a solution
(\ref{puregauge}) which is pure gauge.
These solutions can be easily completed to non linear level when
the source is a perfect fluid. In general this requires both bulk
and brane cosmological constants (the same type of solutions were
also discussed in \cite{csaki} in a different context).
To see how this works recall that in the DGP model the induced
metric on the brane satisfies
\begin{equation}
M_4^2 G_{\mu\nu}-2 M_5^3 (K_{\mu\nu}-g_{\mu\nu} K)=T_{\mu\nu}~,
\label{braneequation}
\end{equation}
where $K_{\mu\nu}$ is the extrinsic curvature on the brane and
$T_{\mu\nu}$ the localized energy momentum tensor. Consistency of
the bulk equation requires that the Gauss equation is satisfied
\cite{sms},
\begin{equation}
G_{\mu\nu}=-\frac {\Lambda_5} {2 M_5^3} g_{\mu\nu}+K
K_{\mu\nu}-K_\mu^\rho K_{\rho\nu} -\frac 1 2
g_{\mu\nu}(K^2-K_{\rho\sigma}K^{\rho\sigma})+E_{\mu\nu}~,
\label{gauss}
\end{equation}
where $E_{\mu\nu}$, the ``dark radiation term'', is traceless and
we have also included a bulk cosmological constant $\Lambda_5$.
The solution in the bulk generically is Schwarzchild-AdS,
$E_{\mu\nu}$ is related to the mass of the bulk black hole (see
Appendix B for details).
Since we look for a solution where the induced metric is flat we
demand,
\begin{equation}
2 M_5^3 (K_{\mu\nu}-\eta_{\mu\nu} K)=-T_{\mu\nu}
\end{equation}
which implies,
\begin{equation}
K_{\mu\nu}=-\frac 1 {2 M_5^3} (T_{\mu\nu}-\frac 1 3 \eta_{\mu\nu}
T).
\end{equation}
The integrability condition (\ref{gauss}) then  determines the
cosmological constant and the dark radiation term. For
$T^\mu_\nu=\rho_0 \, {\rm diag} (-1,~w,~w,~w)$ one obtains
\begin{eqnarray}\label{flat}
E^\mu_\nu&=&-(w+1){\rho_0^2\over 8 M^6_5} \; {\rm diag}
\lp(-1,~\frac 1
3,~\frac 1 3,~\frac 1 3\rp) \nonumber \\
\Lambda_5 &=&(1+3w){\rho_0^2\over 12M^3_5}~.
\end{eqnarray}
Note that $\Lambda_5$ and the black hole mass are quadratic in
perturbation on the brane. As a consequence the energy momentum
tensor appearing in (\ref{kkgraviton}) is only given by the fluid
on the brane to linear level. We postpone the construction of the
full solution to Appendix B. Here we shall advance that for
$w<-1$, the metric takes the form
\beq\label{TaubAds} %
ds_5^2= dy^2-{\sqrt C\over
\ell_{AdS}}{\sinh^2\lp(2(|y|+y_0)/\ell_{AdS}\rp)\over
\cosh\lp(2(|y|+y_0)/\ell_{AdS}\rp)} dT^2 + \sqrt{C}\ell_{AdS}
\cosh\lp(2(|y|+y_0)/\ell_{AdS}\rp) dx_3^2
\eeq %
where
$$
\cosh^2(2y_0/\ell_{AdS})=-{24\over1+w}\lp(\ell_{AdS}\rho_0\over6M_5^3\rp)^2~.
$$
The asymptotic AdS radius is given by
$\ell_{AdS}^2=-{6M^3_5/\Lambda_5}$ and $C$ is an irrelevant scale
with units of length${}^2$.

\subsection{Stability}

An important issue is the stability of the Lorentz violating
backgrounds. Let us first address the ghost instabilities. The
fact that we have a continuous controllable parameter $\rho_0$,
guarantees that there is a finite range for which the solutions
are ghost free. Indeed, since for $\rho_0 = 0$ there are no
ghosts, by continuity this remains true for certain finite range.
The minimal size of this range can be estimated from the following
argument. For very small $\rho_0$, the linearized theory is a good
approximation, and is stable. Ghost instabilities, if they ever
arise, can only set in when the non-linearities become important.
On the other hand, the leading non-linearities come from trilinear
interactions of longitudinal gravitons and their relative strength
is suppressed by the strong coupling scale $\Lambda_{strong}^3 \,
= \,2 M_4/ r_c^2$ \cite{gia}. This gives the most conservative
upper bound on the tension,
\begin{equation}
\rho_0 < \frac {M_4^2} {r_c^2}. \label{lambdamax}
\end{equation}
By taking $r_c$ of the order of the horizon scale this gives
$\rho_0 < (meV)^4$.

The argument above does not rule out the presence of tachyonic
instabilities\footnote{We would like to thank Shinji Mukohyama for
discussions about this point}. These instabilities arise depending
on the fluid and are analogous to the ones of Einstein's static
Universe. Note however that the time scale of these instabilities,
which is under control in the effective description, is set by the
source so it can be made arbitrarily large by appropriate choice
of parameters (see below).

The tachyon instability can be seen from the Friedman equation,
which in DGP with bulk cosmological constant and dark radiation
becomes (see Appendix B),
\begin{equation}
6\epsilon \, M_5^3\, {\sqrt{f(R)+\dot R^2}\over  R} =
\;-3M_4^2\,{\dot R^2+\kappa\over R^2}+{\rho(t)}~, \label{frw}
\end{equation}
where $R(t)$ is the scale factor,
\begin{equation}
f(R)=\kappa - \frac {R^2}{\ell^2} - \frac C {R^2}~,
\end{equation}
$\ell^2={6M^3_5/\Lambda_5}$ is the de-Sitter radius and $C$ is the
mass of the black hole in the bulk. Consider first the case
without DGP term. Eq. (\ref{frw}) implies,
\begin{equation}
\dot{R}^2 - \left( \frac {R^2}{\ell^2} +{C\over R^2}+ \frac
{R^2\,\rho^2} {36 M_5^6}\right)=-\kappa
\end{equation}
This is the equation for a particle with energy $-k/2$ moving in a
potential $V(R)$. From this equation it follows that a negative
bulk cosmological constant (corresponding imaginary $\ell$) can
make the potential positive in some region, and a local minimum
with zero potential is in general possible. This is the case for
some of the solutions that we found. The inclusion of the DGP term
modifies this conclusion in the following way. Any flat space
solution found before is obviously a solution with the DGP term.
If we consider fluctuations around this background, expanding
(\ref{frw}) to quadratic order, $R(t)=R_0+\delta(t)$, we see that
the kinetic term is given by,
\begin{equation}
\left(\frac {3 \epsilon M_5^3}{R_0 \sqrt{f(R_0)}}+ \frac {3 M_4^2}
{R_0^2}\right) \delta'^2
\end{equation}
Using the equation of motion (\ref{frw}) for the solution, one
sees that the kinetic term is
\begin{equation}
3 \frac{6M_5^6+ \rho_0\, M_4^2}{R_0^2\,\rho_0 }\;\delta'^2
\end{equation}
Therefore, if $\rho_0<0$, the DGP term can change the stability of
the solutions making certain unstable configurations stable and
viceversa. In the general with a cosmological constant $\Lambda_4$
and a fluid with energy density $\rho_0'$ and equation of state
$w'$ on the brane, then the second derivative of the potential is
$$
-\frac{(w'+1)\rho_0' \left[(3w'-1)\Lambda_4
+2(3w'+1)\rho_0'\right]}{6M_5^6+[\Lambda_4+\rho_0'] M_4^2}~.
$$
For a generic fluid there is always a range of parameters for
which the solution corresponds to the minimum of the potential and
so it is stable. In the case of instabilities, for realistic
values of the parameters, the time scale of the instability is of
order of Hubble.

\section{De Sitter Breaking}
\label{sec:dS}

In this section we discuss the spontaneous breaking of the
isometries of de-Sitter space. Similar arguments could also be
applied to Anti-de-Sitter.

At the linear level a massive graviton on de-Sitter space must be
described by the PF equation (\ref{pfequation}) where ${\cal E
}^{\alpha\beta}_{\mu\nu}$ is now the linearized Einstein tensor
for de-Sitter space and $\eta_{\mu\nu}$ is replaced by de Sitter
metric,
\begin{equation}
ds^2=-dt^2+e^{2 H t}dx_i dx_i.
\end{equation}
Similarly to the Minkowski case we would like to find solutions
which remain de Sitter space but break its isometries due to the
VEV of the energy-momentum tensor. Following the flat space
example in order to achieve this at linear order one would need to
find solutions of the equation,
\begin{equation}
h_{\mu\nu}-\hat{g}_{\mu\nu} h=-\frac {2\,T_{\mu\nu}} {m^2\,M_4^2}
\label{puregaugeDS}
\end{equation}
which are pure gauge on de Sitter, i.e. $h_{\mu\nu}=D_\mu
\xi_{\nu}+D_\nu \xi_{\mu}$. One can easily prove that this is not
possible at least for any perfect fluid which is homogenous and
isotropic. The crucial point is that due to the cosmological
expansion a conserved $T_{\mu\nu}$ is time dependent,
\begin{equation}
T^\mu_\nu=\frac {\rho_0} {a^{3(1+w)}}\, {\rm diag} (-1,~w,~w,~w),
\end{equation}
where $w$ is the equation of state of the fluid and $a=e^{H t}$.
Solving eq. (\ref{puregaugeDS}) one finds,
\begin{equation}
h^\mu_\nu= \frac 2 {m^2 M_4^2} \frac{\rho_0}{a^{3(1+w)}} \,{\rm
diag} (w+\frac 2 3,-\frac {1} 3 ,-\frac {1} 3,-\frac {1} 3)~.
\end{equation}
Plugging this into the linearized Einstein (or Riemann) tensor one
can easily check that this is not pure gauge for any choice of
$w$. Therefore it is impossible to break Lorentz invariance in
de-Sitter space in a theory with a single massive graviton.

The conclusion above can be circumvented in a theory with several
massive gravitons. The reason for this is that for a general fluid
the solution of the Pauli-Fierz equation on de Sitter is given by,
\begin{equation}
h^\mu_\nu= \frac {2} {(m^2-2 H^2) M_4^2} \frac{\rho_0}{a^{3(1+w)}}
\,{\rm diag} (w+\frac 2 3,-\frac {1} 3 ,-\frac {1} 3,-\frac {1}
3).
\end{equation}
A curios feature of massive gravity on de-Sitter space is that the
only consistent theories must have mass greater than $2H^2$
\cite{higuchi}. It follows that massless and massive gravitons
respond with opposite sign to a fluid. Since the metric measured
is in general a superpositions of modes as in
(\ref{superposition}), it should be possible to arrange a system
so that the total curvature does not change. For example in a
theory with extra-dimensions a fluid localized on a de-Sitter
brane gives rise to a fluid with the same equation of state for
each massive graviton, i.e.
$T_{\mu\nu}^{(m)}=T_{\mu\nu}(0)\psi^{(m)}(0)$ where
$\psi^{(m)}(y)$ is the wave function of the mode $m$. The
five-dimensional metric is then to linear order,
\begin{equation}
h^\mu_\nu(x,y)= \left(\int \frac
{\psi^{(m)}(y)\psi^{(m)}(0)}{2H^2-m^2}\, dm\right) \frac{2\,
\rho_0}{M_4^2\,a^{3(1+w)}} \,{\rm diag} (w+\frac 2 3,-\frac {1} 3
,-\frac {1} 3,-\frac {1} 3). \label{integral}
\end{equation}
It follows that the curvature at the brane is different from zero
unless the integral vanishes due to the cancellation of the zero
mode and the contribution of the entire Kaluza-Klein tower.

Remarkably this possibility can be realized in DGP. To set-up a
de-Sitter background we take $T_{\mu\nu}=-\Lambda_4 g_{\mu\nu}$
and a bulk cosmological constant $\Lambda_5$. As in Minkowski
space to break the isometries we wish to add a non-invariant
energy momentum tensor while keeping the induced metric on the
brane exactly de-Sitter. From eq. (\ref{braneequation}) we see
that this can be achieved by compensating the de-Sitter breaking
energy momentum tensor with the extrinsic curvature,
\begin{equation}
2 M_5^3(K_{\mu\nu}-g_{\mu\nu}K)=-T_{\mu\nu}
\end{equation}
For a general equation of state we have then,
\begin{eqnarray}\label{tensors}
T^\mu_\nu&=&\frac {\rho_0} {a^{3(1+w)}}\; {\rm diag} \lp(-1,~w,~w,~w\rp),\nonumber \\
K_{\nu}^{\mu}&=&\frac 1 {2 M_5^3}\frac {\rho_0} {a^{3(1+w)}}\; {\rm diag} \lp(w+\frac 2 3,-\frac 1 3 ,-\frac 1 3,-\frac 1 3\rp)\nonumber \\
E^\mu_\nu&=&\frac {E_0} {a^4} \; {\rm diag} \lp(-1,~\frac 1
3,~\frac 1 3,~\frac 1 3\rp)~,
\end{eqnarray}
where, since $E^\mu_\nu$ is traceless and conserved, it decays
like radiation.
In order to fulfill the Gauss constraints (\ref{gauss}) the terms
quadratic in $K^\mu_\nu$ should compensate the dark radiation
$E^\mu_\nu$. Given that this scales as $1/a^4$, this requires
$w=-1/3$ . One can then check that Eqs. (\ref{gauss}) are
satisfied by choosing
\beq\label{lambdas} %
\Lambda_4=r_c \Lambda_5~,
\eeq %
and
\beq %
\label{rhorad}%
{E_0} = - {1\over 12} \lp({\rho_0\over M^3}\rp)^2~.
\eeq%
The full bulk solution will be presented in Appendix B. The 5D
geometry is Schwarzchild-de Sitter. In the absence of the fluid,
$\rho_0=E_0=0$,
the bulk is 5D de Sitter and the brane sits at the maximal slice.
This implies that the Hubble rates on the brane and in the bulk
coincide, and that the brane becomes ``stealth''. Indeed, when one
chooses the condition (\ref{lambdas}), the induced gravity term
completely compensates for $\Lambda_4$, and as a result the
effective tension vanishes. Once we turn on the Lorentz violating
fluid on the brane, then the bulk black hole mass is nonzero and
inevitably negative. In order to avoid naked singularities in the
bulk, it turns out that the sign of $\rho_0$ must also be
negative.

Going back to the linearized  solution, despite the presence of
sources in the bulk, the problem still reduces to the one of
massive gravitons coupled to a perfect fluid. This is because, as
can be seen from Eq. (\ref{rhorad}), the sources are second order
in the perturbation, and so they can be neglected in the linear
analysis. What remains to be shown to establish the consistence of
the two solutions is then that the integral (\ref{integral})
indeed vanishes. We prove this in Appendix C.

It is interesting to compare the breaking of the de-Sitter
isometries discussed above with the symmetry breaking due to the
condensation of higher spin fields. Consider for example the
spin-1 tachyon. Since the stabilizing potential is a function of
$A_\mu A^\mu$ it is clear that a solution with a constant VEV of
$A_0$ is not affected by the cosmological expansion. This is
possible because the tachyon condensate does not reduce to an
ordinary fluid. In the DGP case since the symmetry breaking is
induced by a conserved energy momentum tensor on the brane the
violations of isometries dilutes with the expansion of the
universe.

\section{Conclusions}
\label{conclusions}

In this note we investigated different mechanisms for breaking
spontaneously Lorentz invariance or more in general isometries of
maximally symmetric spaces. One possibility is the condensation of
integer spin tensor fields. This scenario in general implies the
presence of ghost instabilities which cannot be controlled in an
effective field theory framework around a maximally symmetric
background. The second possibility is breaking Lorentz invariance
through the VEV of the energy momentum tensor. In ordinary four
dimensional theories this would necessarily affect the geometry of
space-time leading to practically unobservable Lorentz violating
effects once cosmological bounds are enforced. This conclusion
does not hold in a theory with extra-dimensions such as DGP or
even in a theory with massive gravitons because the space can
remain maximally symmetric despite the Lorentz violation.

\section*{Acknowledgments}

We would like to thank Nima Arkani-Hamed, Jose Juan Blanco-Pillado
and especially Gregory Gabadadze and Shinji Mukohyama for
interesting discussions. The work of G. D. and M. R. is supported
in part by David and Lucile Packard Foundation Fellowship for
Science and Engineering, and by the NSF grant PHY-0245068. O. P.
acknowledges support from Departament d'Universitats, Recerca i
Societat de la Informaci{\'o} of the Generalitat de Catalunya,
under the Beatriu de Pin{\' o}s Fellowship 2005 BP-A 10131.

\section*{Appendix A: Arbitrary Spins}

In this appendix we extend the results of section \ref{hs} to higher
spin fields. A massive spin$-n$ field can be decomposed into a
massless spin-$n$ and St\"uckelberg fields of lower spin. For a
spin-$n$ an entire tower of massless spin-$(n-1),...,0$ St\"uckelberg
fields will be required. The most convenient way to construct
consistent theories of high spin massive particles is by starting
with the massless theory in one extra-dimension \cite{riccioni}. A
massless spin$-n$ field is described by a symmetric tensor
$H_{\mu_1\dots\mu_n}$ subject to a ``double trace'' constraint,
\begin{equation}
H^{\alpha\beta}_{~~\alpha\beta\,\mu_5\dots\mu_n}=0.
\label{doubletrace}
\end{equation}
As for the spin-2, at the quadratic level, there is a unique
ghost-free consistent lagrangian,
\begin{eqnarray}
{\cal L}&=&-\frac 1 2 (\partial_M H_{\dots})^2+\frac n 2 (\partial_M
H^M_{~\dots})^2+\frac {n(n-1)} 2 (\partial_M H^L_{~L\dots})^2
\nonumber \\
&& + \frac {n(n-1)(n-2)} 8 (\partial_M H^{ML}_{~~L\dots})^2+\frac
{n(n-1)} 2 H^L_{~L\dots}(\partial_M\partial_N H^{MN}_{~~\dots})^2
 \label{highspin}
\end{eqnarray}
which is gauge invariant with respect to,
\begin{equation}
\delta H_{M_1\dots M_n}=n\,\partial_{(M_1} \epsilon_{M_2\dots M_n)}
\label{agauge}
\end{equation}
where $\epsilon_{M_2\dots M_n}$ is a traceless symmetric tensor and
the parenthesis denotes symmetrization of the indexes. The massive
theory in one dimension less can be derived by compactifying on a
circle of radius $1/m$ and truncating to the first KK level.  One of
the advantages of this formulation is that it keeps the gauge
invariance of the massive theory manifest. In fact the components of
the field with indexes in the internal dimension and harmonic
dependence on the extra coordinate are automatically the
St\"uckelberg fields of the theory,
\begin{equation}
H_{\mu_1\dots\mu_s\,y\dots y}(x,y)=\phi^{(s)}_{\mu_1\dots \mu_s}(x)
\cos (m\,y), \label{astuckelberg}
\end{equation}
From (\ref{agauge}) it follows that the massive theory is
invariant under,
\begin{equation}
\delta \phi^{(s)}_{\mu_1\dots \mu_s}= s\, \partial_{(\mu_1}
\epsilon_{\mu_2\dots \mu_s\,y\dots y)}-(n-s) m\,
\epsilon_{\mu_1\dots\,\mu_s\,y\dots y}, \label{shift}
\end{equation}
Roughly speaking the meaning of this transformation is that each
$\phi^{(s)}$ is a gauge field with an additional shift symmetry
required to to be Goldstone boson for $\phi^{(s+1)}$. Differently
from the spin-2 case however not all the St\"uckelberg fields can be
gauged away and some of them act as auxiliary fields (this follows
from the constraint that $\epsilon$ must be traceless).

Let us now focus on the highest spin St\"uckelberg field
$\phi^{(n-1)}_{\mu_1\dots\mu_{n-1}}$. By plugging
(\ref{astuckelberg}) into the action (\ref{highspin}) one can see
that this field acquires a positive kinetic term of the form
(\ref{highspin}). Note also that no mass term is generated because
of the global shift transformation in (\ref{shift}). For the lower
St\"uckelberg fields the situation is more complicated in general
because the ones that are dynamical acquire kinetic terms from the
mixing with higher ones.

The previous derivation applies to the positive $m^2$ spin$-n$
theories. The tachyonic case can be derived by starting with two
time directions and compactifying one of them. This reverses the
sign of all the mass terms. The kinetic term for the first
St\"uckelberg is now ghost-like proving that any higher spin tachyon
contains ghost instabilities. The lower St\"uckelbergs
which are dynamical will have alternating positive and ghost kinetic
terms.

\section*{Appendix B: Exact Solutions}

In this Appendix we present the full solutions including the bulk
where the brane is maximally symmetric even though it supports
Lorentz violating sources. We shall focus on two cases: in the
first one, the brane is flat and the source is a perfect fluid
with arbitrary equation of state $w=p/\rho$. The Weyl tensor is
non-zero, and the bulk is asymptotically $AdS$.
In the second example, the brane geometry is de Sitter and the
source is a fluid with equation of state $w=-1/3$. As we shall
see, this solution requires a positive bulk cosmological constant
$\Lambda_5$ and a non-zero Weyl tensor as well.

In order to construct the solution, we note that a maximally
symmetric  brane is compatible with a bulk whose symmetries are
only the isotropy and homogeneity of the 3-dimensional spatial
slices, {\em i.e.}, 5D spherical symmetry (see also the discussion
in \cite{csaki}). Given that in the bulk we have Einstein gravity,
the Birkhoff theorem ensures that the bulk takes the form of the
Schwarzchild-de Sitter metric \cite{bcg},
\beq\label{static} %
ds_5^2= -f(R) dT^2 +{dR^2\over f(R)} +R^2 d\Omega_\kappa^2
\eeq %
where $f(R)=\kappa - (R/\ell)^2 - C/R^2$, and $d\Omega_\kappa^2$
denotes the line element of a unit 3 dimensional sphere, flat
space or hyperbolic space for $\kappa=1,~0$ and $-1$ respectively.
The de Sitter curvature radius is given by
$$
\ell^2={6M_5^3\over \Lambda_5}~,
$$
and the projected Weyl tensor \cite{sms} (see Eq (\ref{tensors}))
is related to the 5D Black hole mass parameter $C$ through
\beq\label{CrhoE} %
E_0 = {3 C\over R^4}~.
\eeq %

The brane location can always be parameterized by two functions
$R=R(t)$ and $T=T(t)$. This introduces a reparameterization
freedom, that can be fixed by choosing
\beq\label{gchoice} %
f(R)\, \dot T^2 -{\dot R^2 \over f(R)} =1~.
\eeq %
In this gauge, the induced metric on the brane is simply
\beq\label{induced} %
ds_4^2=-dt^2+R^2(t) d\Omega_\kappa^2.
\eeq %
The brane trajectory, that is the form of $R(t)$, is obtained from
the Israel junction condition (\ref{braneequation}). This reduces
to the modified Friedman equation
\beq %
\label{angular}%
6\epsilon \, M_5^3\, {\sqrt{f(R)+\dot R^2}\over  R} =
\;-3M_4^2\,{\dot R^2+\kappa\over R^2}+{\rho(t)+\Lambda_4}~.
\eeq%
Here, $\rho(t)$ is the energy density of the fluid on the brane,
and  $\epsilon=\pm1$ is the sign of the extrinsic curvature. In
our conventions, $\epsilon=1$ means that the bulk is the
``interior'' of the brane.

\subsection*{B.1 Flat Brane with General Fluid}

In this case, the full solution can be read off from
(\ref{static}). In the flat slicing ($\kappa=0$), the metric has a
horizon at $R=(C\ell^{2})^{1/4}$, where now $\ell$ denotes the AdS
radius, $\ell^2=-{6M^3_5/\Lambda_5}$. The brane sits at a constant
``radial'' coordinate, $R_0$. Its actual value (together with $C$)
is not determined by the Friedman equation (\ref{angular}) or by
(\ref{gauss}), which only determine the ratio $E_0={3 C/ R_0^4}$.
What this means is that we can trade a larger black hole with a
brane location closer to it. However, once we specify the distance
between the brane and the horizon, everything else is fixed.
In terms of the proper coordinate along the bulk the metric takes
the form presented in (\ref{TaubAds}). In those coordinates, it is
apparent that the parameter $C$ is irrelevant, as it can be
rescaled away. The constant $y_0$ is the distance between the
brane and the horizon, which is entirely fixed by $\rho_0$ and
$w$. Note also that (\ref{TaubAds}) explicitly goes to flat space
in the $\rho_0\to0$ limit.

>From Eq. (\ref{flat}), we see that for $w<-1$, the bulk black hole
has positive mass, and it also follows that $\Lambda_5<0$. Given
that the brane is flat, a fluid with this equation of state can be
composed of the brane cosmological constant $\Lambda_4$ plus some
fluid with energy density $\rho_0'$ and equation of state $w'$.
For $w'>-1$ and $\Lambda_4<-\rho_0'$, one obtains that the
equation of state of the superposition is $w<-1$. This also
implies that the total energy density on the brane is negative. In
Figure \ref{fig:diagrams}, we show the conformal diagram
corresponding to this case. The brane lies outside the black hole
horizon, and the bulk corresponds to the region that includes the
asymptotic AdS boundary.

For $w=-1/3$ the metric reduces to the one found by Taub long ago
\cite{taub},
\begin{equation}
ds^2= \lp(1+c|y|\rp)^{-{2\over 3}} \lp( -dt^2 + dy^2 \rp)+
\lp(1+c|y|\rp)^{2\over 3} dx_{3}^2 %
\label{taub}
\end{equation}
with $c=-\rho_0/M^{3}_5$. As in the solutions of the next
subsection, the energy density of the brane $\rho_0$ must be
negative in order not to have a naked singularity in the bulk.

\subsection*{B.2 De Sitter Brane}

Let us now turn to the solutions with a de Sitter brane. Since the
induced metric is given by (\ref{induced}), we will have a de
Sitter brane by demanding that $R(t)$ is
\beq %
R(t)= H^{-1} \left\{
\begin{array}{ll}
    \cosh(H t) & \hbox{for~$\kappa=1$} \\
    e^{Ht} & \hbox{for~$\kappa=0$} \\
    \sinh(H t) & \hbox{for~$\kappa=-1$}~. \\
\end{array}%
\right.     %
\eeq%
Let us now find out what kind of fluid gives rise to this
trajectory. The scale factor solves the equation
$$
\dot R^2+\kappa=(HR)^2~.
$$
If the brane and bulk expansion rates coincide, $H=1/\ell$, the
extrinsic curvature
$$
{1\over R}\sqrt{f(R)+\dot R^2}={\sqrt{-C}\over R^2}
$$
contributes to (\ref{angular}) like a curvature term. In light of
this, the Friedmann equation (\ref{angular}) reduces to
\beq %
\label{angular2}%
6\epsilon \, M_5^3\,{\sqrt{-C}\over R^2} = -3M_4^2\,H^2+\Lambda_4
+\rho~.
\eeq%
Hence, it is possible to preserve the de Sitter expansion if
$H^2=\Lambda_4/3M_4^2$ and the fluid redshifts like $1/R^2$, which
corresponds to an equation of state $w=-1/3$.
If the brane and the bulk inflate at different rates, then a de
Sitter brane is still in principle possible, but this requires a
fluid with a very specific time dependent equation of state.
Hence, we shall disregard this possibility. Let us emphasize that
the condition $H=1/\ell$ is equivalent to
$$
\Lambda_4=r_c \Lambda_5,~
$$
and that in the absence of the fluid (and the bulk black hole) it
implies that the brane is effectively \emph{stealth}: for this
choice of $\Lambda_4$, the brane can only 'fit' in the equator of
the 5D de Sitter, so the extrinsic curvature must vanish. This is
compatible with the fact that the brane has nonzero tension thanks
to the DGP term, which completely screens it, so that the
effective tension is zero.

It is clear from (\ref{angular2}) that the bulk black hole mass
parameter $C$ can only be negative. This means that there is a
naked singularity at $R=0$. Thus, the only regular solution must
include the 'exterior' of the brane. This corresponds to negative
extrinsic curvature, $\epsilon=-1$. Hence, the fluid must also
have $\rho<0$.

It is not apparent in the static coordinates (\ref{static}), but
the extension of the Schwarzschild-de Sitter metric beyond the
cosmological horizon has another naked singularity placed at the
opposite pole, as shown in Fig. \ref{fig:diagrams}. To avoid it,
one can place another brane with the same $\Lambda_4$ and  with
the same kind of fluid with $w=-1/3$.

Once $R(t)$ is fixed, $T(t)$ is obtained by Eq. (\ref{gchoice}).
After some algebra, one arrives to the following equation for
$T(R)$,
\beq\label{T(R)} %
\partial_R T = {\sqrt{-C}\over R f(R) \sqrt{(HR)^2-\kappa}}~.
\eeq %
This also tells us that the only possible dS embedding is with a
negative mass particle in the bulk, $C<0$.

Let us briefly comment on the continuation beyond the cosmological
horizon. For any $\kappa$, $R(t)$ eventually grows without bound.
The static coordinates displayed in (\ref{static}) in general
cover a finite range of $R$ bounded by the two roots of $f(R)=0$.
In our case ($C<0$ and $\Lambda_5>0$) there is no inner horizon
and the outer horizon is at $R=R_+>\ell$. Hence, the brane
initially fits inside the horizon but in a finite proper time it
crosses it.
The continuation of (\ref{static}) is done by replacing $T\to
X+i\pi/2$ and simply allowing for $R>R_+$, so that the metric
looks like $ds_5^2= g(R) dX^2 -dR^2/ g(R) +R^2 d\Omega_\kappa^2 $
with $g(R)=-f(R)>0$.
Thus, $R$ becomes the time coordinate and it continues to be the
scale factor on the brane. Given that this continuation does not
affect the junction condition (\ref{angular}), the form of $R(t)$
of course does not change when $R_+$ is crossed.
We can also see that  $T$ picks up a constant imaginary part from
Eq. (\ref{T(R)}), since the r.h.s. has a simple pole at $R=R_+$.

\begin{figure}[!tb]
\begin{center}
\includegraphics[width=.7 \textwidth]{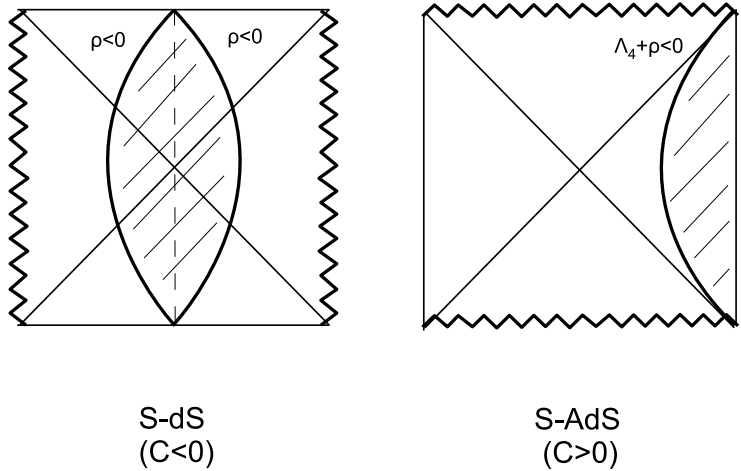}
\caption{Left: conformal diagrams of the solution with de Sitter
branes. The bulk consists of Schwarzschild-de Sitter with negative
mass, which has two naked singularities, one at each pole. The
static chart displayed in (\ref{static}) covers the interior of
the left diamond. In order to avoid the singularity, we have to
take $\rho<0$. Still, we have to avoid the other one, which can be
accomplished by placing another  brane with the same kind of fluid
after the equator, represented with the dashed line. Right:
conformal diagram for the solution with a flat brane supporting a
fluid with generic equation of state. The bulk is
Schwarzschild-AdS with positive mass. The effective brane tension
is negative, so the solution corresponds to the exterior,
including the AdS boundary. The background solution does not
depend on $M_4$, but gravity is not effectively four dimensional
unless the DGP term is included.}
\label{fig:diagrams} %
\end{center}
\end{figure}

\section*{Appendix C: Linearized Analysis in de Sitter}

In this appendix, we complete the linearized analysis of the de
Sitter violating solution discussed in Section \ref{sec:dS}.
Recall that for the de Sitter brane solution, the background bulk
spacetime is five dimensional de Sitter space,
$$
ds^2=a^2(y)ds_4^2+dy^2~,
$$
with $a(y)=H^{-1}\cos(H y)$.
Introducing a Kaluza-Klein decomposition $h_{\mu\nu}=a^2(y)\sum_m
\psi_m(y)h^{(m)}_{\mu\nu}(x)$ where the 4D modes obey
(\ref{pfequation}), then one find the following equation for the
wave functions,
$$
\lp[ a^{-4}\partial_y a^4 \partial_y + {m^2\over a^2}+ r_c m^2
\delta(y)\rp] \psi_{m}(y)=0~.
$$
This equation determines that the spectrum is composed of a
massless mode and a continuum of massive gravitons starting at
$m^2=9/4\,H^2$. Instead of $m$, it is convenient to label each KK
level by $p$, defined as
$$
m^2=H^2\lp(p^2+9/4\rp)~.
$$
Thus, the continuum of KK modes starts at $p=0$, while the zero
mode is at $p=3i/2$. Plugging the KK decomposition in the equation
of motion, we obtain an equation for the wave functions
$\psi_m(y)$ whose general solution is a linear combination of
$$
U_{ip}(y)= {\Gamma(1-ip)\over \sqrt{2\pi}\,2^{ip}H^{-3/2}} \,{P_{i
p}^{3/2}\lp(\sin(Hy)\rp)\over \cos^{3/2}(Hy)}~,
$$
and its conjugate $U_{-ip}$, which are the analogues of the plane
waves in de Sitter. Here, $P_\nu^\mu(z)$ denotes the Legendre
function of the first kind and the constants are chosen so that
they are normalized to a delta function.
$$
\int_{-\pi/2H}^{\pi/2H} dy\, a^2(y)\,(1+r_c\delta(y))\,
U_{ip}(y)U_{-ip'}(y)=\delta(p-p')~.
$$
The $r_c\delta(y)$ term in this normalization is required because
of the kinetic term localized on the brane (see \cite{dgpfluct}
for a detailed discussion).

The precise combination is dictated by the boundary conditions,
which take the form
$$
\partial_y \psi - {1\over2} r_c H^2 (p^2+9/4) \psi=0~,
$$
where everything is evaluated at $y=0^{+}$. Hence, KK wave
functions are
\beq\label{KK} %
\psi^{KK}_p(y)={A(-p) U_{ip}-A(p) U_{-ip}\over i \sqrt{2
A(p)A(-p)}}~,
\eeq %
where
$$
A(p)= U'_{ip}(0) - {1\over2} r_c H^2 (p^2+9/4) U_{ip}(0)~.
$$
The zero mode is normalized as
$$
\int_{-\pi/2H}^{\pi/2H} dy\, a^2(y)\,(1+r_c\delta(y))\,
|\psi_{z.m.}(y)|^2=1~.
$$
Given that $\psi_{z.m.}(y)$ is constant, we obtain
$$
\psi_{z.m.}^2=\lp({\pi\over2}+Hr_c\rp)H^3~.
$$

In order to show that to linear order the induced metric remains
de Sitter, we need to compute the sum over the spectrum appearing
in Eq. (\ref{integral}). This can be written as
$$
I=-{1\over 2 H^2 }\psi_{z.m.}(0)^2+{1\over H^2 }\int_0^\infty dp\,
{{\psi^{KK}_p(0)}^2\over p^2+1/4}~.
$$
The contribution from the KK modes can be expressed as
\beq\label{int} %
{1\over2}\int_{-\infty}^{\infty}dp \,{1\over p^2+1/4}\lp(U_{ip}
U_{-ip}-{A(-p) U_{ip}^2\over A(p)}\rp)
\eeq %
This can be converted into contour integral where the contour
closes the upper half plane. The integral, then, is evaluated by
summing over the residues. Given that the Legendre functions have
no poles in the $p$ plane, the only poles contributing to
(\ref{int}) come from the zeros of $A(p)$ and possibly from
$p=i/2$. It is easy to show that in the upper $p$ plane $A(p)$
vanishes at $p=3i/2$. A straightforward computation of the residue
shows that this contribution completely cancels that of the zero
mode. On the other hand, when the wave functions are evaluated on
the brane, at $y=0$, the residue at $p=i/2$ turns out to be zero.
Hence,  $I=0$.

\vspace{0.5cm}



\begin{thebibliography}{99}

\bibitem{coleman}
  S.~R.~Coleman and S.~L.~Glashow,
  ``High-energy tests of Lorentz invariance,''
  Phys.\ Rev.\  D {\bf 59}, 116008 (1999)
  [arXiv:hep-ph/9812418].

\bibitem{papucci}
  G.~Dvali, M.~Papucci and M.~D.~Schwartz,
  ``Infrared Lorentz violation and slowly instantaneous electricity,''
  Phys.\ Rev.\ Lett.\  {\bf 94}, 191602 (2005)
  [arXiv:hep-th/0501157].

\bibitem{arkani}
  N.~Arkani-Hamed, H.~C.~Cheng, M.~A.~Luty and S.~Mukohyama,
  ``Ghost condensation and a consistent infrared modification of gravity,''
  JHEP {\bf 0405}, 074 (2004)
  [arXiv:hep-th/0312099].

\bibitem{dgp}
  G.~R.~Dvali, G.~Gabadadze and M.~Porrati,
  ``4D gravity on a brane in 5D Minkowski space,''
  Phys.\ Lett.\ B {\bf 485}, 208 (2000)
  [arXiv:hep-th/0005016].

\bibitem{luty}
  H.~C.~Cheng, M.~A.~Luty, S.~Mukohyama and J.~Thaler,
  ``Spontaneous Lorentz breaking at high energies,''
  JHEP {\bf 0605}, 076 (2006)
  [arXiv:hep-th/0603010].

\bibitem{extratime}
  G.~Dvali, G.~Gabadadze and G.~Senjanovic, 1999 unpublished. See also
  G.~Gabadadze, Proceedings of ``Cosmo 99'', Trieste, Italy, 1999.

\bibitem{lue}
  C.~Deffayet, G.~R.~Dvali, G.~Gabadadze and A.~Lue,
  ``Braneworld flattening by a cosmological constant,''
  Phys.\ Rev.\ D {\bf 64}, 104002 (2001)
  [arXiv:hep-th/0104201].

\bibitem{oriol}
  G.~Dvali, G.~Gabadadze, O.~Pujolas and R.~Rahman,
  ``Domain walls as probes of gravity,''
  [arXiv:hep-th/0612016].

\bibitem{gia}
  C.~Deffayet, G.~R.~Dvali, G.~Gabadadze and A.~I.~Vainshtein,
  ``Nonperturbative continuity in graviton mass versus perturbative
  discontinuity,''
  Phys.\ Rev.\  D {\bf 65}, 044026 (2002)
  [arXiv:hep-th/0106001];
  M.~A.~Luty, M.~Porrati and R.~Rattazzi,
  ``Strong interactions and stability in the DGP model,''
  JHEP {\bf 0309}, 029 (2003)
  [arXiv:hep-th/0303116].

\bibitem{csaki}
  C.~Csaki, J.~Erlich and C.~Grojean,
  ``Gravitational Lorentz violations and adjustment of the cosmological
  constant in asymmetrically warped spacetimes,''
  Nucl.\ Phys.\  B {\bf 604}, 312 (2001)
  [arXiv:hep-th/0012143].

\bibitem{sms}
  T.~Shiromizu, K.~i.~Maeda and M.~Sasaki,
  Phys.\ Rev.\  D {\bf 62}, 024012 (2000)
  [arXiv:gr-qc/9910076].

\bibitem{higuchi}
  A.~Higuchi,
  ``Forbidden mass range for spin-2 field theory in de Sitter space-time,''
  Nucl.\ Phys.\  B {\bf 282}, 397 (1987).

\bibitem{riccioni}
  M.~Bianchi, P.~J.~Heslop and F.~Riccioni,
  ``More on la grande bouffe,''
  JHEP {\bf 0508}, 088 (2005)
  [arXiv:hep-th/0504156].

\bibitem{bcg}
  P.~Bowcock, C.~Charmousis and R.~Gregory,
  ``General brane cosmologies and their global spacetime structure,''
  Class.\ Quant.\ Grav.\  {\bf 17} (2000) 4745
  [arXiv:hep-th/0007177].

\bibitem{taub}
  A.~H.~Taub,
  ``Plane symmetric spacetimes,''
  Phys.\ Rev.\  {\bf 103}, 454 (1956).

\bibitem{dgpfluct}
  O.~Pujolas,
  ``Quantum fluctuations in the DGP model and the size of the cross-over
  scale,''
  JCAP {\bf 0610}, 004 (2006)
  [arXiv:hep-th/0605257].


\end{thebibliography}
\end{document}